# Miniaturized Shack-Hartmann Wavefront-Sensors for Starbugs

Michael Goodwin*[a], Samuel Richards[a,b], Jessica Zheng[a], Jon Lawrence[a], Sergio Leon-Saval[b], Alexander Argyros[b], Belen Alcalde[a]

[a]Australian Astronomical Observatory, PO Box 915, North Ryde, NSW 1670, Australia; [b]School of Physics, The University of Sydney NSW 2006, Australia

## ABSTRACT

The ability to position multiple miniaturized wavefront sensors precisely over large focal surfaces are advantageous to multi-object adaptive optics. The Australian Astronomical Observatory (AAO) has prototyped a compact and lightweight Shack-Hartmann wavefront-sensor that fits into a standard Starbug parallel fibre positioning robot. Each device makes use of a polymer coherent fibre imaging bundle to relay an image produced by a microlens array placed at the telescope focal plane to a re-imaging camera mounted elsewhere. The advantages of the polymer fibre bundle are its high-fill factor, high-throughput, low weight, and relatively low cost. Multiple devices can also be multiplexed to a single low-noise camera for cost efficiencies per wavefront sensor. The use of fibre bundles also opens the possibility of applications such as telescope field acquisition, guiding, and seeing monitors to be positioned by Starbugs. We present the design aspects, simulations and laboratory test results.

**Keywords:** Starbugs, Shack-Hartmann, Miniaturized, Wavefront-Sensors, Adaptive Optics

## 1. INTRODUCTION

Adaptive Optics (AO) [1] is an essential facility for large ground-based telescopes to correct the aberrations induced by atmospheric turbulence to maximize telescope efficiency. Recent techniques such as Multi-Object Adaptive Optics (MOAO) [2] have increased the need for multiple wavefront sensors placed at the focal plane. Positioning of these devices to within the accuracy and time constraints is a challenging problem. To tackle this problem we propose the concept of miniature shack-hartmann wavefront sensors (mini-SHWFS) positioned with a standard 'Starbugs' [3, 4] developed by the Australian Astronomical Observatory (AAO). Starbugs provide the ability to simultaneous walk fibres around the focal with accuracies better than 5 microns and configuration times better than several minutes. Starbugs are a new type of fibre positioner that has demonstrated well both in the laboratory and early on-sky tests. Starbugs are proposed for the TAIPAN [5] and MANIFEST [6, 7] instruments.

Wavefront sensing has a multitude of applications in Astronomy, particularly that of Adaptive Optics (AO) [1]. The purpose of AO is to remove the optical aberrations from the optical path from the science object and the imaging detector to improve the image quality. Nearly all of the wavefront aberrations are induced by the atmospheric turbulence as random phase perturbations along the line-of-sight path. The wavefront sensor attempts to measure these phase perturbations in real-time, typical on timescales of milliseconds (where turbulence is assumed to be 'frozen'). The wavefront sensor can also measure telescope tracking errors and wind buffeting, as well as slow timescale aberrations such as mirror/dome seeing and mirror gravity distortions. The low frequency errors (spatial and temporal) are the largest errors and are controlled by a separate system known as "active optics".

There are number of AO modes of correction (field size and image quality) to achieve different science goals. The correction various from the highest level of correction (full correction) to the lowest correction (partial correction). The highest levels of correction require the most accurate wavefront measurements, such as Extreme AO (ExAO) [8]. The opposite is the case for the lowest levels of correction, such as Ground Layer AO (GLAO) [9]. It is difficult to correct over large fields sizes and hence the image quality also tends declines (partial corrections). The exception being the more advanced techniques, such as Multi-conjugate AO (MCAO) [10] and Multi-object AO (MOAO). The application we propose for our mini-SHWFS are those requiring a large number of mid-to-low wavefront measurements (for partial corrections), such as GLAO and MOAO.

---

* michael.goodwin@aao.gov.au; http://www.aao.gov.au; phone +61 2 9372 4851; fax +61 2 9372 4860

## 2. CONCEPT

The concept of mini-SHWFS positioned by Starbugs is shown in Figure 1. The use of polymer coherent imaging bundles allows the relay of the image formed by each Starbug WFS to multiplex onto a single fast readout, low noise detector. The same concept can also be applied to miniature curvature wavefront sensors also proposed by the AAO [11] but using two coherent imaging bundles to simultaneously relay the intra- and extra- focus images from a telescope.

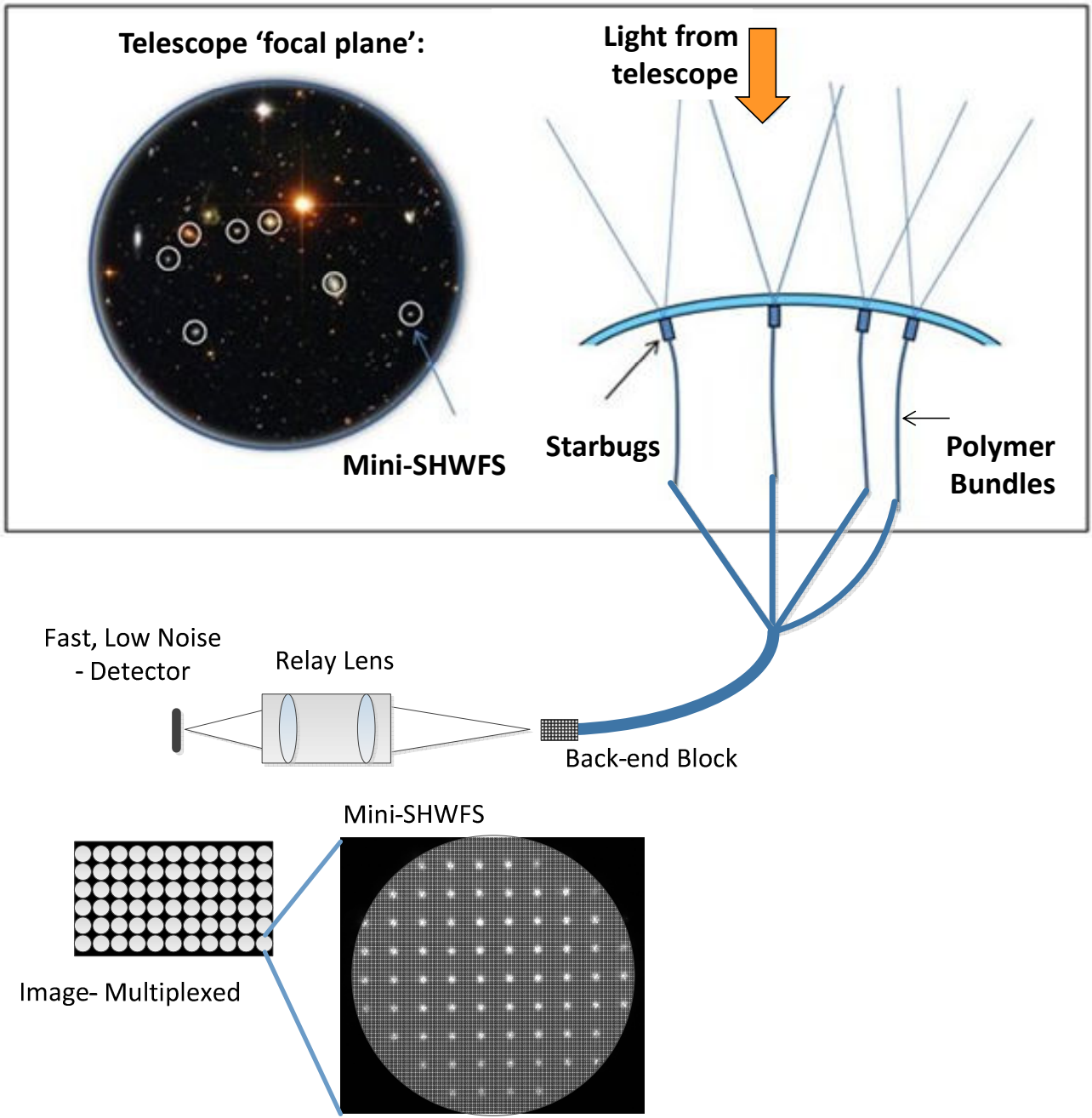

Figure 1: Concept diagram for mini-SHWFS using Starbugs.

The operational concept is as follows: Starbugs are first positioned on a suitable guide star (natural or laser) and are held onto the glass field plate by a vacuum. Each Starbug has fore-optics consisting of a tiny collimating lens that images the telescope pupil onto a microlens array (MLA). The MLA array then images its spot pattern onto the polymer coherent imaging bundle. The bundles from multiple Starbugs are collated and anchored to form a single back-end block. A relay lens then re-images the back-end block (with suitable magnification) onto a fast, low noise detector. The detector image consists of a multiplexed array of mini-SHWS images. The software then processes the detector image to reconstruct the individual wavefronts of each mini-SHWS. Flexible operation modes such as higher framerates or improved sensitivity (limiting magnitude of guide star) can be built into the system to adapt to the observing conditions. This can be achieved by taking advantage of detector region of interest (ROI) reads, detector binning and dynamic range (11-bit, 16-bit). The multiplex of multiple mini-SHWFS onto a single fast, low noise detector (most expensive component) is a significant cost savings to the instrument.

## 3. DESIGN

### 3.1 Starbug Devices

Starbugs [3, 4] are the next concept in parallel robotic positioning to position optical fibres precisely at the focal plane to catch the light from objects and relay them through optical fibres to stable bench-mounted spectrographs. The AAO has proposed Starbugs as the fiber-positioner for TAIPAN [5] and MANIFEST [6, 7] instruments. The goal of Starbugs is to reconfigure typical fields in a several minutes that are typically the downtime between fields to either move the telescope or read the detector. Starbugs are adaptable to their focal surfaces and can re-position over flat or curved focal surfaces. An individual Starbug, see Figure 2 (a), comprises two piezoceramic tube actuators, joined at one end to form a pair of concentric 'legs' that can be electrically driven to produce a micro-stepping motion in the ±x and ±y directions (i.e. forwards, backwards, left, right) and θ (rotation). Discrete step sizes typically ranging from 3 to 20 microns at 100 Hz provide precise positioning with speeds up to few millimeters per second. Starbugs are positioned under closed-loop control and are monitored with a camera imaging their back-illuminated metrology fibres, see Figure 2 (b). The outer diameter of a Starbug is approximately 8 mm with tube length approx. 20 mm, providing capability to carry payloads with diameter of 3.75 mm, see Figure 2 (b). Tests have shown that Starbugs can successfully position 1.5 mm polymer coherent imaging bundles on the UKST at Siding Spring Observatory, Australia [5].

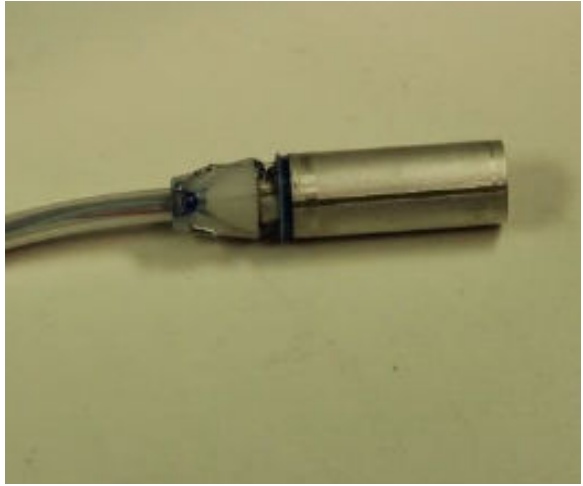 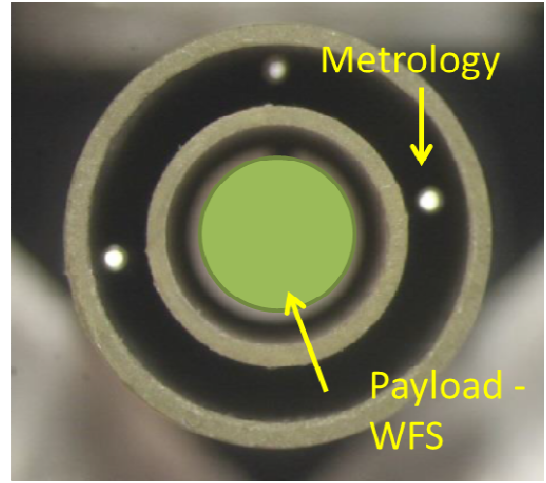

(a) Exterior diagram of the Starbug with a tube length of approximately 20 mm and diameter of 8 mm.

(b) Starbug front surface (contacts with glass field plate to form a vacuum) showing the inner and outer tubes, the metrology fibres and the location for the payload (diameter ~ 3.75 mm).

Figure 2: Starbug mechanical (a) and nominal payload position (b).

### 3.2 Polymer coherent imaging bundles

Polymer coherent imaging bundles are used to relay the image of each mini-SHWFS to the detector. These bundles are 1.5 mm in diameter and contain 7000 cores that are each approx. 16 microns in diameter. The efficiency has been measured to be approximately 75% with a numerical aperture (NA) ranging from 0.16 to 0.37, allowing the relay of focal beams as fast as F/1.3 to 3.1 [11]. The AAO has used the polymer coherent imaging bundles for the SAMI instrument [12, 13] as the guide bundle module relaying the image from 3 guide stars (22" field of view each) to the existing guide camera (with minimal software changes). The installation of the SAMI guide bundle module has eliminated previous guide camera flexure and allowed extra room to configure the plug-plate without compromising the brightness of the guide source (limiting magnitude). Given the success of using polymer coherent imaging bundles for SAMI, they are proposed for the TAIPAN guide bundle module, to provide functions of field plate alignment, field acquisition, telescope guiding and as a plate reference for positioning. A TAIPAN prototype consisting of 7 guide bundles is shown in Figure 3 as tested on the 1.2 m UKST telescope. The uniformity in the flat illumination is excellent with minimal defects. Tests on the 1.2 m UKST telescope (early tests: proto-TAIPAN [5]) provided acquisition of V~10 magnitude stars with exposures approx. 40 ms (25 fps).

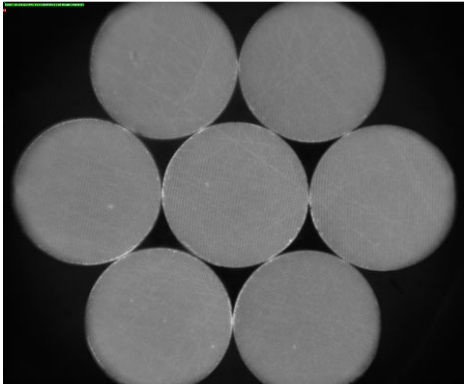
(a) flat field illumination

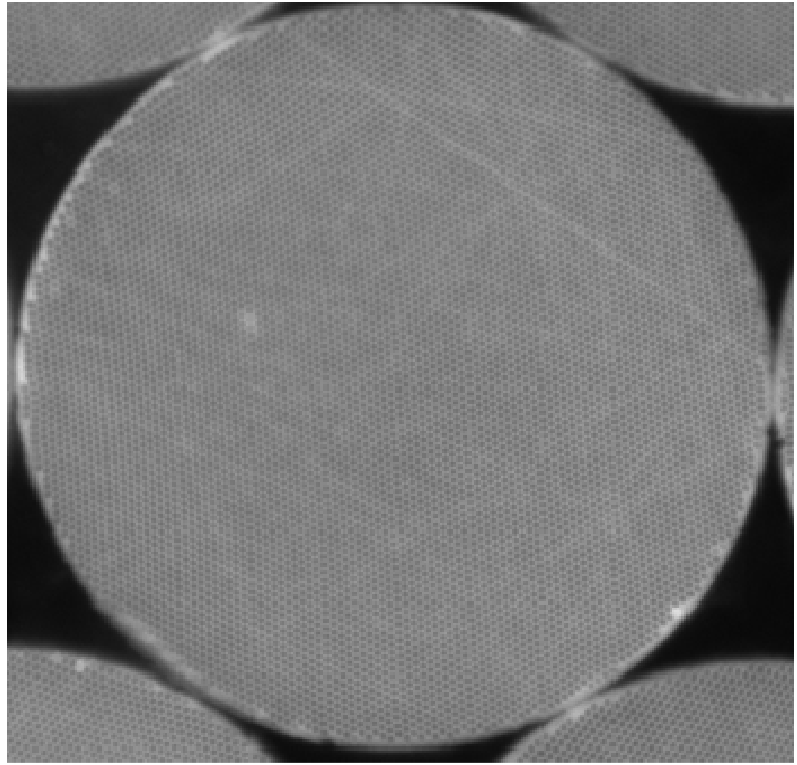

(b) close-up of a single bundle as shown in (a)

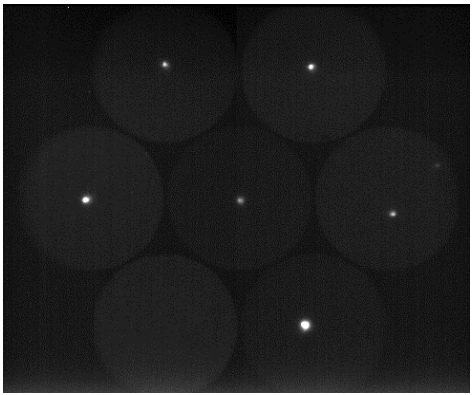
(c) imaging V~10 stars with 40 ms exposures

Figure 3: Polymer coherent bundles used for early TAIPAN tests on the UKST 1.2 m. Each bundle is 1.5 mm or 100" field.

## 3.3 Starbugs payload (fore-optics)

The fore-optics design for the mini-SHWFS for the Anglo-Australian Telescope (AAT) Cassegrain focal f/8 beam is shown in Figure 4. The fore-optics is designed in ZEMAX with off-the-shelf components. The components are miniaturized to fit into the allowable dimensions of the Starbug. The collimating lens being 3 mm in diameter with the length from the telescope focus to bundle being approx. 29 mm. The MLA array selected to provide adequate number of sub-apertures over the AAT 3.9 m pupil, being 13x13, or sub-aperture length of approx. 30 cm, see Figure 5. Increasing the number of sub-apertures causes spot confusion, particularly after sampled by the bundle cores and in bad seeing conditions. The mini-SHWFS is designed for dynamic range of +/- 1.8", which is sufficient for most conditions (telescope guiding and tracking are usually the largest errors). Note it is important to take into consideration that the purpose for these mini-SHWFS is for partial corrections – not precise WFS, so tolerances can be somewhat relaxed.

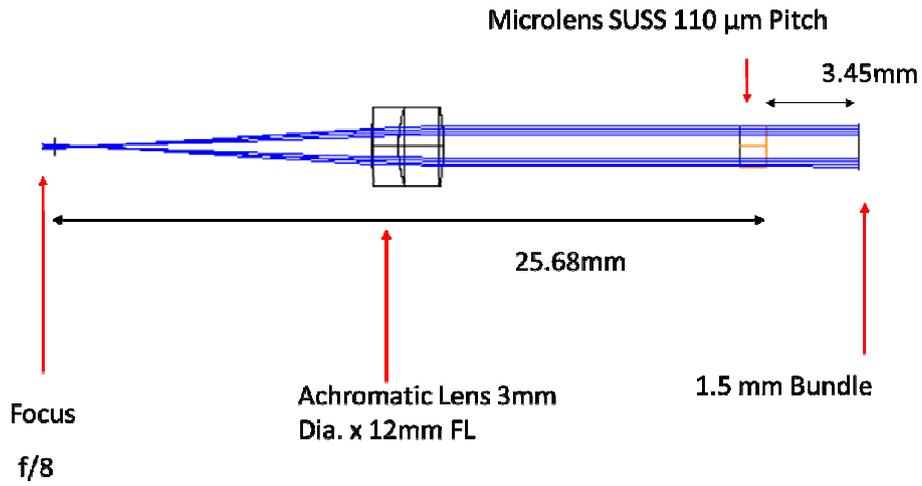

Figure 4: ZEMAX optical design for mini-SHWFS for the AAT f/8 that fits into a standard Starbug.

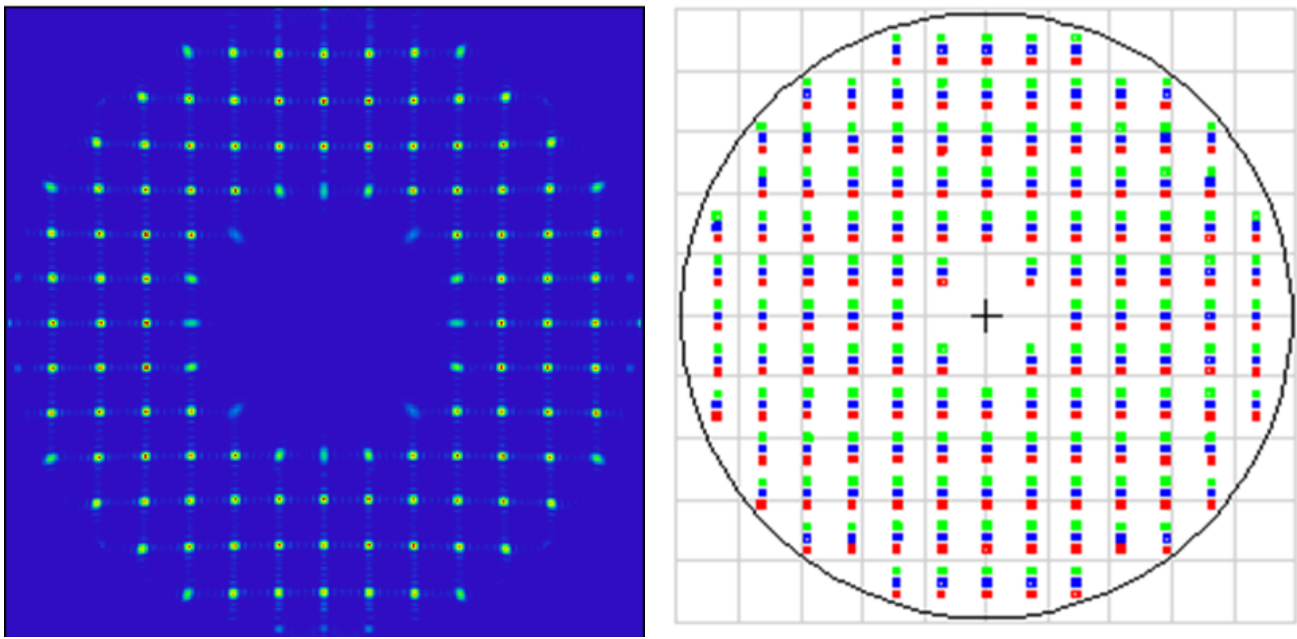

(a) ZEMAX Physical Optics Propogation 13x13 Sub-apertures 30 cm

(b) ZEMAX footprint diagram +/- 0.0005 deg or 1.8"

Figure 5: ZEMAX diagrams showing the spot patterns for the mini-SHWFS for Starbugs. Note that the core-sampling of the polymer bundle has not been simulated in ZEMAX.

## 3.4 Back-end re-imaging module

The back-end module re-images the back-end block of the polymer bundle array onto the detector, as shown in Figure 6. The design is similar to that used for the SAMI and TAIPAN guide bundle modules. To achieve the correct pixel sampling a relay lens is used with an appropiate magnification. Camera binning can be used to achieve the correct sampling but improvement in sensitivity is marginal if the camera has low read noise values. For the design in Figure 6, an aluminium block is used as the frame that provides a sturdy support to components. The frame is designed interface to a c-mount camera, Thorlabs SM1 relay lens and the bundle array back-end block, that can be interchanged.

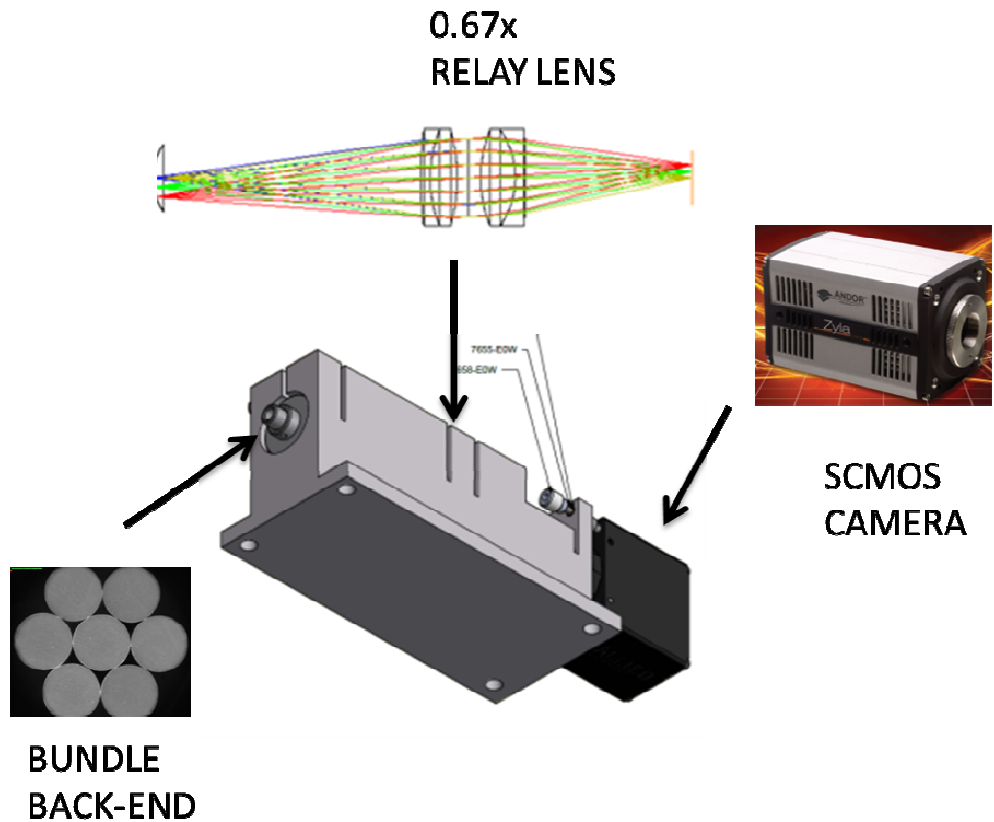

Figure 6: Back-end module concept design to re-image the bundles onto the detector.

Using the Andor Zlya DG-152X-CIE-F1 sCMOS camera [14] the back-end re-imaging module has two planned modes – fast (419) and slow (198) frame reads using ROI frame reads, see Figure 7 and Table 1. Although it is possible ot have more modes depending on the ROI. However the downside is that increased frame rates comprimises the detector area (number of bundles imaged). The Andor Zlya sCMOS is capable of 5 M pixel reads at 100 fps with a read noise 1.45-1.80 electrons with a rolling shutter [14].

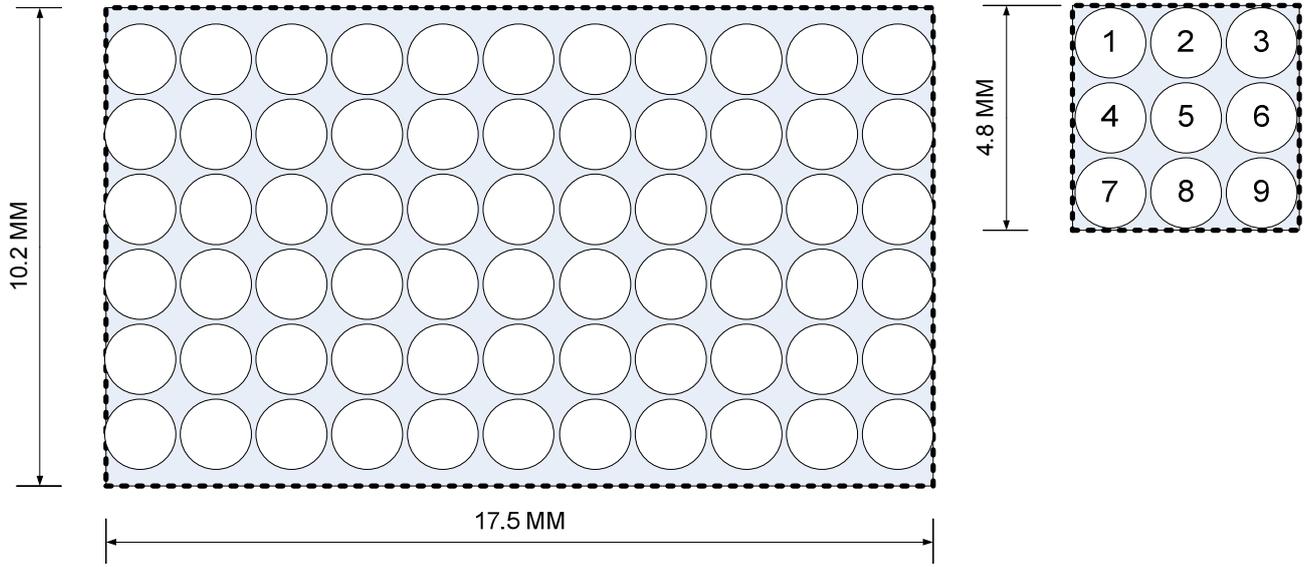

Figure 7: Number of mini-SHWFS imaged on the back-end bundle block for fast (left) and slow (right) modes.

Table 1: Description of fast and slow modes for the back-end re-imaging module.

| Mode | ROI (pixels) | Object ROI (mm) | Image ROI (mm) | FPS | Bundles | SHWFS Dim. |
|---|---|---|---|---|---|---|
| FAST | 512x512 | 4.8x4.8 | 3.328 x3.328 | 419 | 3x3=9 | 13x13 (110um MLA) |
| SLOW | 1920x1080 | 10.2x17.5 | 12.480 x7.020 | 198 | 6x11=66 | 13x13 (110um MLA) |

## 4. SIMULATIONS

A numerical simulation of the polymer coherent image bundle based mini-SHWFS serves as part of the design verification process. Hence a model was constructed using MATLAB to simulate a single 13x13 sub-aperture SHWFS detector images obtained with and without using the coherent imaging bundle. The SHWFS being simulated for the AAT 3.9 m at f/8 Cassegrain focus as a component of the conceptual 66 bundle mini-SHWFS running at 198 fps. The system overview is shown in Figure 8. The simulated detector images are first processed for spot locations (centroid frame - local wavefront gradients) and then processed to reconstruct the wavefront (Zernike polynomials). The reconstructed wavefront is then compared with the reference Kolmogorov wavefront, in terms of residual root mean squared (RMS) and point spread function (PSF).

The diagram showing the steps involved for simulating the detector images for the mini-SHWFS and classical SHWFS are shown in Figure 9. The simulation only considers a single fixed Kolmogorov screen located at the telescope pupil, height of 0 km (does not simulate scintillation or layer wind speed). The simulation parameters are listed in Table 2. Illustration of the sampling of the sub-aperture PSFs onto the bundle for a Flat (locations for centroid regions) and Kolmogorov wavefronts is shown in Figure 10. Enlargement of the right of mid-section of the simulated SHWFS detector images with centroid data overly is shown in Figure 11.

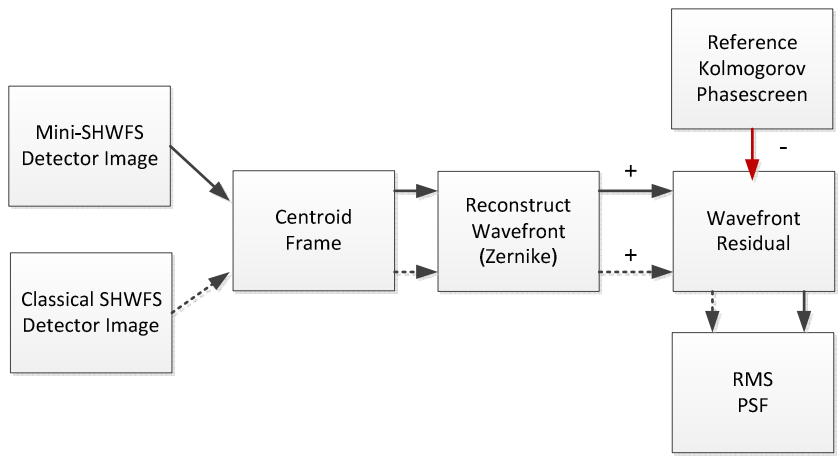

Figure 8: System overview of the mini-SHWFS simulation model.

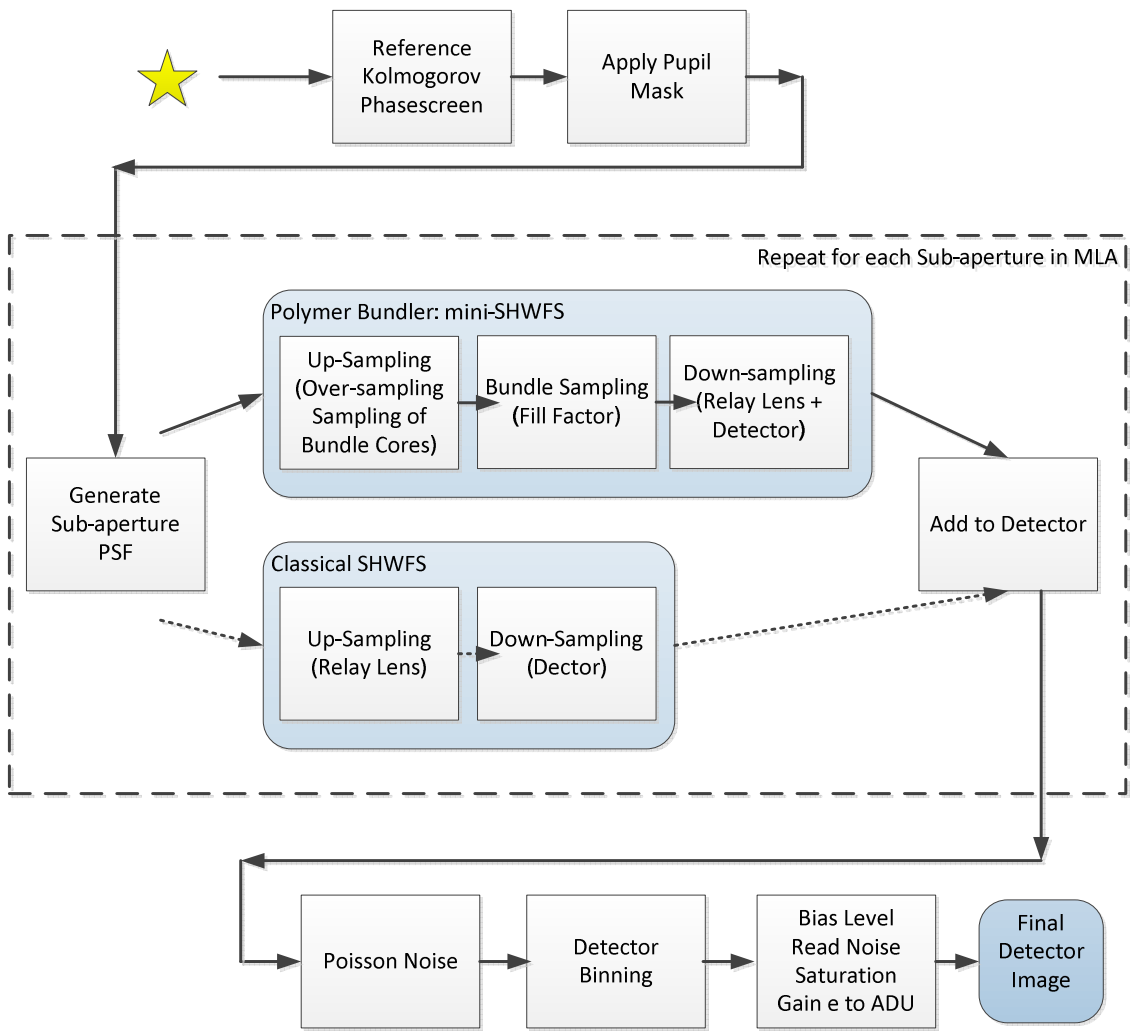

Figure 9: Process flow diagram that describes the generation of simulated detector images.

Table 2: Simulation parameters for the SHWFS

| Parameter | Value |
|---|---|
| SHWFS Dimension | 13 x 13 sub-apertures |
| Pupil Diameter (D) | 3.9 m (AAT) |
| Pupil Obstruction Ratio | 0.4 D |
| SHWFS Wavelength | 0.5 microns (measurement) V-band |
| Science Wavelength | 1.25 microns (correction) J-band |
| Simulation Exposure | 5 ms (note 1) |
| Simulation Star Magnitudes | 5, 9, 10, 11 (V-band) |
| Efficiency | 0.25 (source to detector) |
| Wavefront Layer | 1 (0 km, Kolmogorov, no wind) |
| Wavefront Coherent Length, $r_0$ | 0.1 m (1 arcsec, V-band) (note 2) |
| Wavefront Sampling (Pupil) | 0.0125 m / pixel |
| Wavefront Reconstruction | Zernike Polynomials – 45 Terms |
| Bundle Diameter | 1.5 mm |
| Bundle Core Diameter | 14.5 microns |
| Bundle Core Pitch | 17 microns |
| Bundle Core Sampling | 1 pixel = 0.975 microns |
| MLA Pitch | 110 microns (0.3 m sub-apertures) |
| MLA Focus Length | 3.45 mm |
| Relay Lens Magnification | 0.67 |
| Detector Type | Andor Zlya DG-152X-CIE-F1 sCMOS (note 3) |
| Detector Pixel Size | 6.5 microns |
| Detector Read Noise | 1.45 electrons |
| Detector Bias Level | 50 ADUs |
| Detector ADU Gain | 0.6 electrons / ADU |
| Detector Saturation | 30000 electrons |
| Detector Binning | 1 x |
| Detector ROI | 1920 x 1080 (6x11 = 66 bundles) |
| Detector FPS | 198 (Rolling Shutter) |

Notes: (1) Exposure time 5 ms the layer can be consider 'frozen' over sub-aperture: timescale~0.3*d/v, [1] where d is the sub-aperture width and v is average wind speed, in our case d=0.3 m and v=10 m/s, timescale~9 ms. Also, maximum exposure time is limited by camera frame rates of 198 fps.

(2) The atmospheric seeing of 1 arcsec (note: mirror/dome component removed) is the best 30% of seeing conditions measured at Siding Spring Observatory, Australia [15].

(3) Detector parameters are taken from the manufacture's datasheet [14].

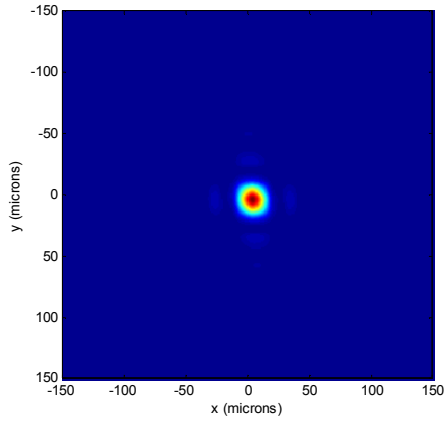
(a) Bundle front-end

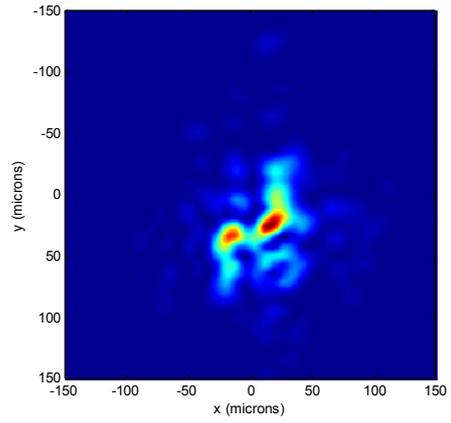
(b) Bundle front-end

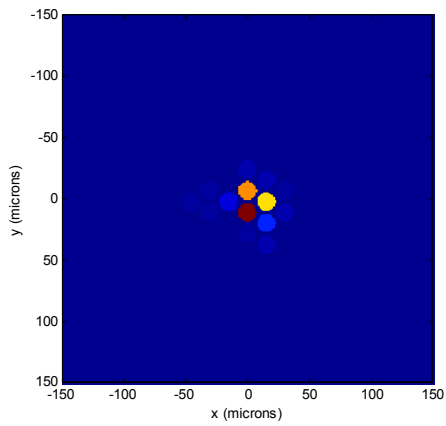
(c) Bundle back-end

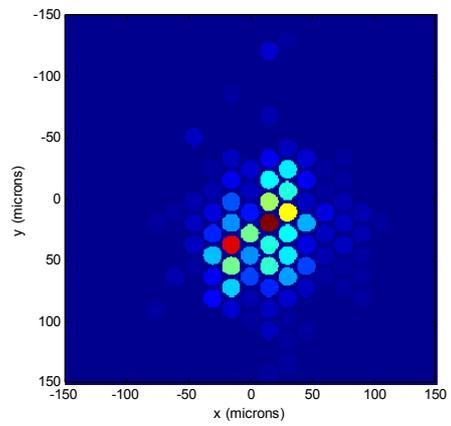
(d) Bundle back-end

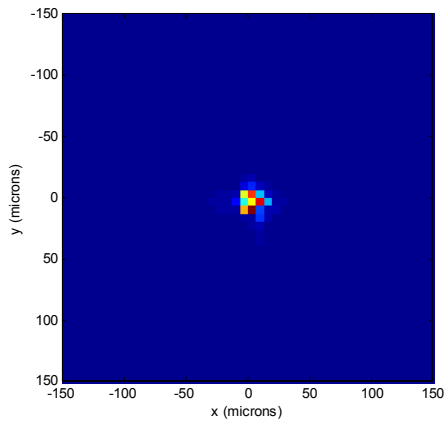
(e) Detector (after Relay-Lens)

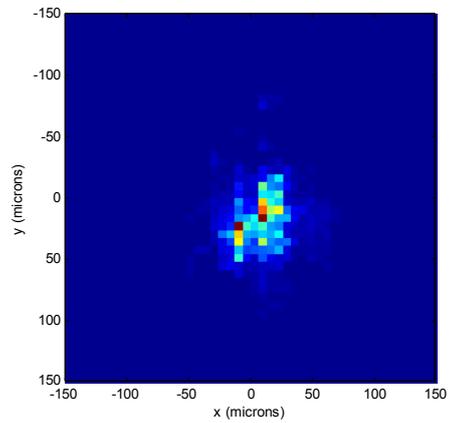
(f) Detector (after Relay-Lens)

Figure 10: Sub-aperture PSF sampling onto the detector (see Figure 9) for a Flat wavefront (a), (c), (e); and for a Kolmogorov wavefront (b), (d), (f). Bundle back-end images (c) and (d) are highly-oversampled to model the circular cores and fill factor. The images (c) and (d) are magnified by 0.67 by the image relay lens and then binned to detector pixel resolution PSF images (e) and (f).

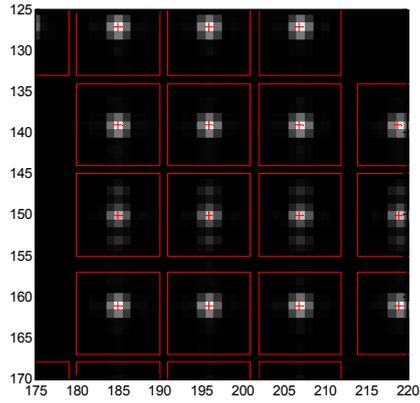
(a) Classical SHWFS (Flat wavefront)

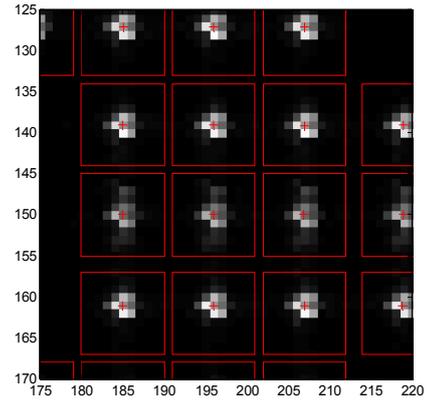
(b) Bundle mini-SHWFS (Flat wavefront)

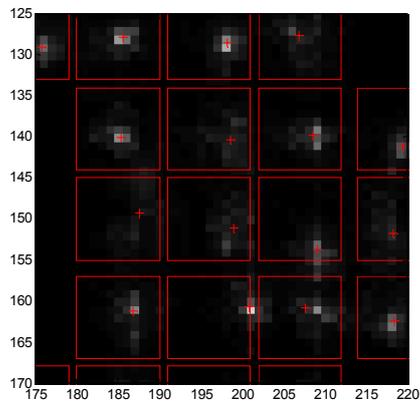
(c) V-band=5

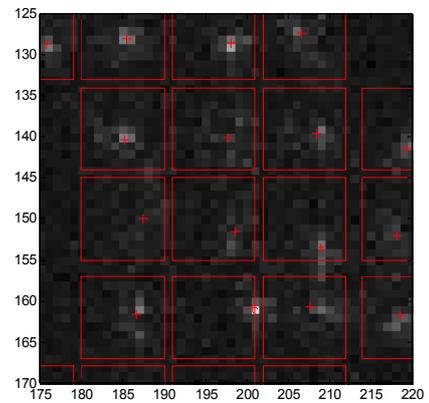
(d) V-band=9

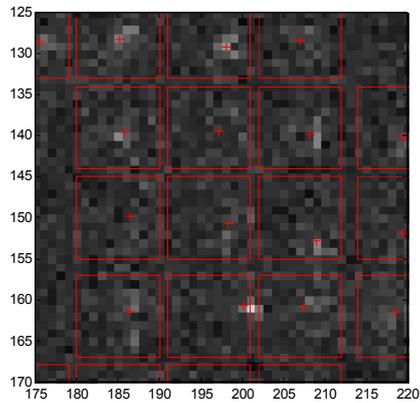
(e) V-band=10

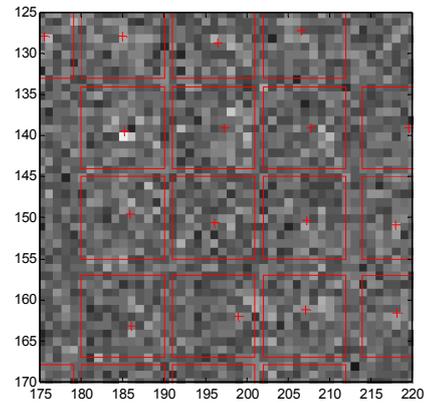
(f) V-band=11

Figure 11: Enlargement of the right of mid-section of the simulated 13x13 SHWFS detector images with centroid location overlays (boxes) with centroid positions (marked with '+'). Images (a) and (b) are the corresponding flat wavefront images used to calculate centroid regions. Images (c) to (f) are simulated guide stars of increasing faintness for the bundle mini-SHWFS, exposure is 5 ms. Each image is scaled according to detector maximum.

From the Flat wavefront shown in Figure 11 (a) and (b) we see that bundle sampling blurs the PSF. The fine details of the bundle sampled detector PSF (Figure 10 (f)) in Figure 11 (c) to (f) are not shown due to image contrast. An example result for the bundle mini-SHWFS simulation is shown in Figure 12 and shows the improvement in PSF (J-band) based on the residual wavefront error (measured minus the reference wavefront). The result in Figure 12 only takes into consideration only the spatial sampling (13x13 sub-apertures over 3.9 m pupil) and bundle sampling only, ignoring other errors that may be present in the AO system. The low spatial sampling is assumed to be the largest error component for the intended applications of our mini-SHWFS. The RMS residual for the classical and mini-SHWFS for V-band magnitudes is listed in Table 3. The 'slope gain' referred in Table 3 is the multiplication factor for the detector spot displacements or sub-aperture slopes (Kolmogorov wavefront minus the Flat wavefront).

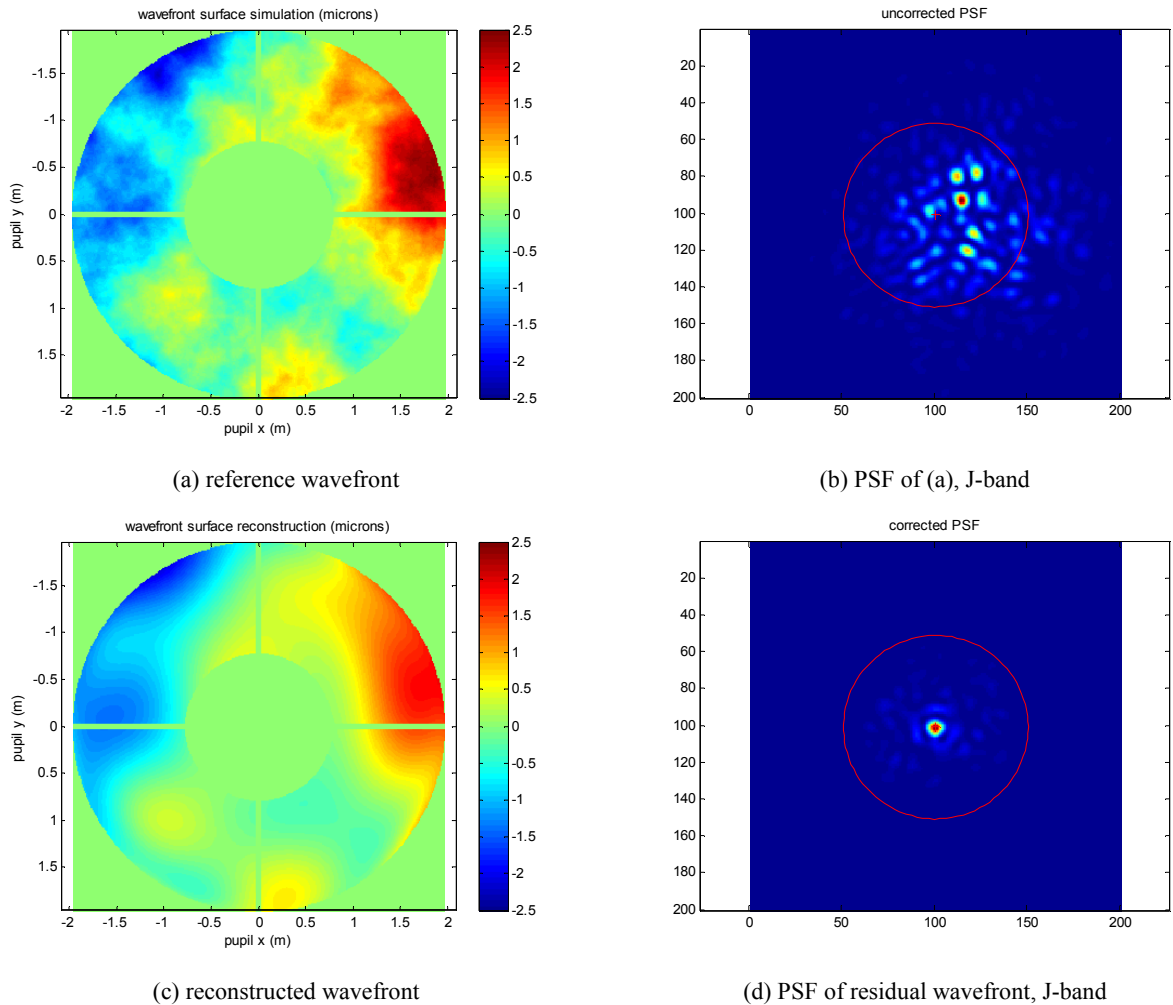

(a) reference wavefront  (b) PSF of (a), J-band

(c) reconstructed wavefront  (d) PSF of residual wavefront, J-band

Figure 12: Simulation results for the bundle mini-SHWFS for a guide star V-band=9 and exposure 5 ms in 1 arcsecond seeing (slope gain =1) achieves a residual wavefront RMS of 0.2116 microns. The circle in (b) and (d) has a diameter of 1 arcsecond.

Table 3: The RMS of the residual wavefront in microns for SHWFS with guide star V-band magnitudes 5 to 11, slope gain=1. The RMS of the Kolmogorov wavefront reference is 0.8459 (microns).

| SHWFS \ V-mag | 5 | 9 | 10 | 11 |
|---|---|---|---|---|
| Classical (RMS) | 0.1980 | 0.2242 | 0.2945 | 0.4902 |
| Polymer bundle (RMS) | 0.1984 | 0.2116 | 0.3172 | 0.5417 |

The 'slope gain' is need when the noise dominates the signal, as shown in Figure 11 (f), as the centroid is biased towards the central region. This causes a decrease in the wavefront peak-valley of the reconstructed wavefront, as illustrated in Figure 13 (a). By increasing the slope gain, as in Figure 13 (b), the amplitude is near that of the simulated wavefront, Figure 13 (a). The decrease in peak-valley of the wavefront when the signal-noise-ratio drops is also observed with the classical SHWFS. The improvement by 'tuning' the slope gain to minimize the residual wavefront RMS (to achieve the best image quality) for the classical SHWFS is listed in Table 4 and the bundle mini-SHWFS listed in Table 5. The improvements are slight, but clearly noticeable for V-band=11 guide star. The slope gain increases with V-band magnitude (fainter guide stars) and are higher for the bundle mini-SHWFS.

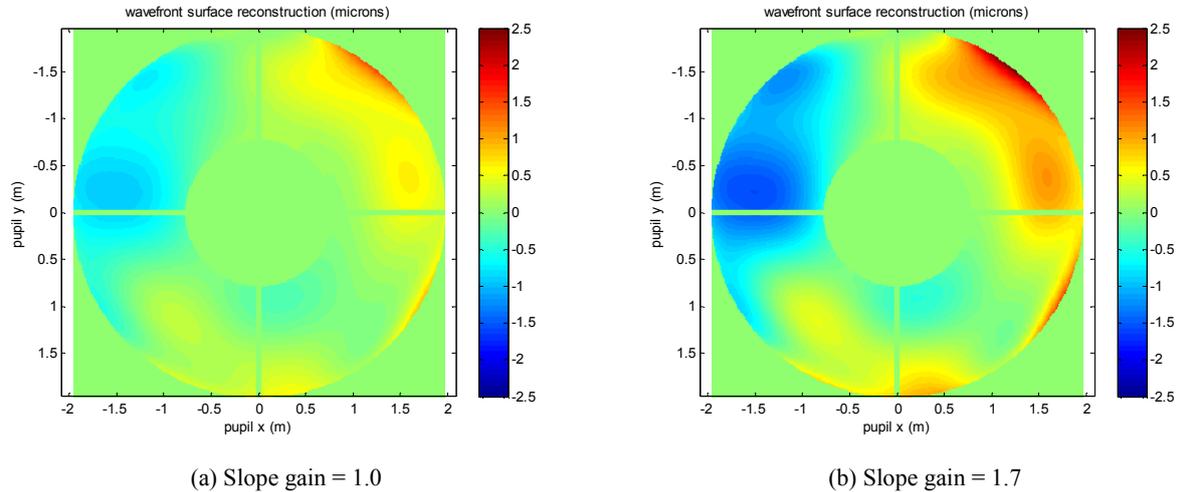

(a) Slope gain = 1.0  (b) Slope gain = 1.7

Figure 13: The reconstructed wavefront for polymer bundle mini-SHWFS for V-band=11. The slope gain is adjusted to minimise the RMS of the residual wavefront. When the noise dominates the signal, the reconstructed wavefront has a lower peak-valley (a) that can be compensated or 'tuned' by increasing the slope gain as in (b).

Table 4: The RMS of the residual wavefront in microns for classical SHWFS with guide star V-band magnitudes 5 to 11, slope gain adjusted to minimise the RMS. The RMS of the Kolmogorov wavefront reference is 0.8459 (microns).

| SHWFS \ V-mag | 5 | 9 | 10 | 11 |
|---|---|---|---|---|
| Classical (RMS) | 0.1980 | 0.2141 | 0.2454 | 0.3976 |
| Slope gain | 1.00 | 1.09 | 1.3 | 1.5 |

Table 5: The RMS of the residual wavefront in microns for polymer bundle mini-SHWFS with guide star V-band magnitudes 5 to 11, centroid gain adjusted to minimise the RMS. The RMS of the Kolmogorov wavefront reference is 0.8459 (microns).

| SHWFS \ V-mag | 5 | 9 | 10 | 11 |
|---|---|---|---|---|
| Polymer bundle (RMS) | 0.1984 | 0.2031 | 0.2308 | 0.4562 |
| Slope gain | 1.00 | 1.05 | 1.4 | 1.7 |

## 5. EXPERIMENTAL RESULTS

The experimental setup to study the feasibility of using polymer coherent image bundles for mini-SHWFS is shown in Figure 14. The experiment was conducted as part of a 12-week AAO Summer Student Fellowship project. The experiment uses a Thorlabs AOKit to generate the wavefronts specified in Zernike terms on a 144 actuator deformable mirror (DM) to all direct comparison of the measurements obtained with the bundle mini-SHWFS (located at the beam dump) with that of the Thorlabs SHWFS, see Figure 14. Two bundles were available for testing, a 2.5 mm diameter bundle with 26 micron separation between cores and the other 1.5 mm bundle with 17 micron separation between cores. Example detector images of a Flat wavefront are shown in Figure 15 using a 150 micron pitch MLA.

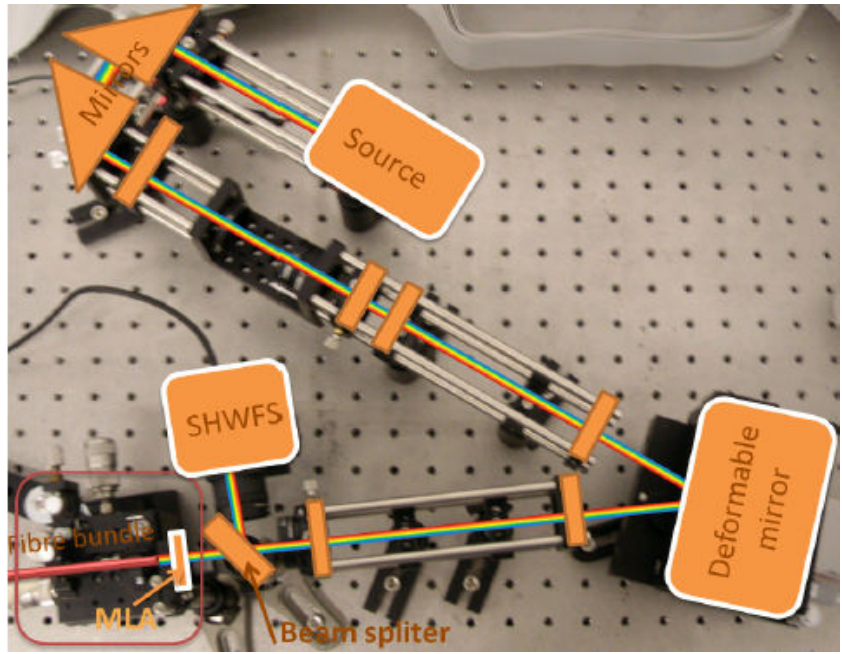

Figure 14: Experimental setup to compare the measured wavefronts using a polymer bundle mini-SHWFS (bottom left corner) with that of a Thorlabs SHWFS using the Thorlabs Adaptive Optics kit.

The results of the 2.5 mm bundle are listed in Table 6 and the 1.5 mm bundle in Table 7. The RMS of the residual wavefronts for some cases is better than $\lambda/4$ but others are the RMS is much higher. Through experiment, it was found RMS errors to be sensitive to external factors such alignments (pupil on MLA between the two WFS) and contact with components, etc. The tip-tilt usually contributed a significant fraction of the RMS error. An improvement to the experiment is to integrate the mini-SHWFS into a single package rather than separate components. An example result showing good agreement between the bundle mini-SHWFS and Thorlabs SHWFS is shown in Figure 16.

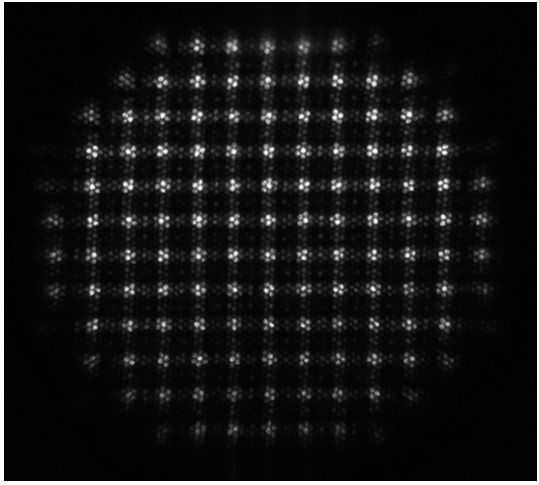 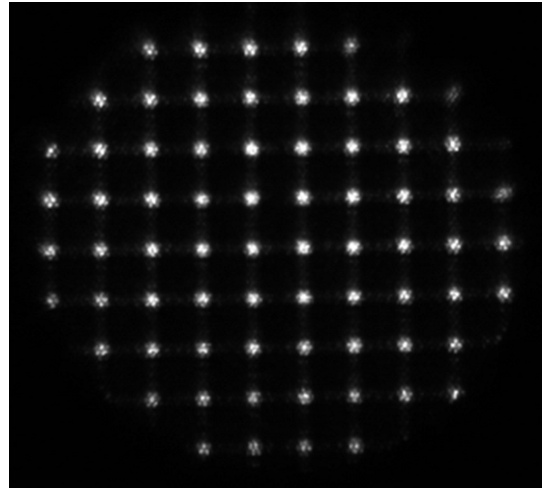

(a) 2.5 mm bundle with 150 micron MLA. Bundle core separation approx. 26 microns.

(b) 1.5 mm bundle with 150 micron MLA. Bundle core separation approx. 17 microns.

Figure 15: Reference images (flat wavefronts) of the bundle mini-SHWFS using the experimental setup shown in Figure 15.

Table 6: Summary of the laboratory measurements (Zernike terms & error characterization) results comparing 150 micron MLA with the 2.5 mm diameter polymer fibre bundle WFS (FBWFS) with the Thorlabs SHWFS. The last 4 columns are specified in microns.

| Aberration | | Zterm | value | FBWFS | Th.SHWFS | $WF\Delta$ | medfit | RMS | PV |
|---|---|---|---|---|---|---|---|---|---|
| Tilt | Vert. | z2 | 0.3 | -0.189 | -0.249 | 0.060 | 0.025 | 0.091 | 0.434 |
| | | | 0.5 | -0.393 | -0.416 | 0.023 | 0.025 | 0.050 | 0.256 |
| | Horizt. | z3 | 0.3 | 0.364 | 0.248 | 0.116 | 0.012 | 0.215 | 0.852 |
| | | | 0.5 | 0.515 | 0.426 | 0.089 | 0.001 | 0.135 | 0.535 |
| Defocus | | z4 | 0.3 | -0.217 | -0.236 | 0.019 | 0.008 | 0.999 | 0.514 |
| | | | 0.5 | -0.465 | -0.484 | 0.020 | -0.020 | 0.213 | 0.897 |
| Astigmatism | Oblique | z5 | 0.3 | 0.283 | 0.303 | 0.020 | 0.027 | 0.216 | 1.336 |
| | | | 0.5 | 0.424 | 0.462 | 0.038 | 0.010 | 0.312 | 1.594 |
| | Vert. | z6 | 0.3 | 0.244 | 0.283 | 0.039 | -0.017 | 0.114 | 0.590 |
| | | | 0.5 | 0.412 | 0.460 | 0.047 | 0.010 | 0.196 | 1.109 |
| Coma | Vertical | z7 | 0.3 | -0.202 | -0.239 | 0.038 | -0.100 | 0.159 | 0.760 |
| | | | 0.5 | -0.308 | -0.264 | 0.044 | -0.031 | 0.239 | 1.193 |
| | Horizt. | z8 | 0.3 | -0.184 | -0.212 | 0.028 | 0.036 | 0.172 | 1.196 |
| | | | 0.5 | -0.226 | -0.225 | 0.001 | 0.083 | 0.213 | 1.173 |
| Trefoil | Vert. | z9 | 0.3 | 0.250 | 0.286 | 0.036 | -0.348 | 0.249 | 1.151 |
| | | | 0.5 | 0.286 | 0.176 | 0.110 | -0.250 | 0.238 | 1.255 |
| | Oblique | z10 | 0.3 | 0.136 | 0.140 | 0.004 | 0.014 | 0.034 | 0.241 |
| | | | 0.5 | 0.207 | 0.229 | 0.021 | -0.002 | 0.007 | 0.341 |
| z3,z5 & z7 | | z3 | 0.2 | 0.253 | 0.131 | 0.122 | | | |
| | | z5 | 0.2 | 0.189 | 0.225 | 0.036 | 0.018 | 0.148 | 0.976 |
| | | z7 | 0.2 | -0.118 | -0.120 | 0.002 | | | |
| | | z3 | 0.3 | 0.253 | 0.113 | 0.140 | | | |
| | | z5 | 0.3 | 0.224 | 0.289 | 0.065 | -0.001 | 0.192 | 1.111 |
| | | z7 | 0.3 | -0.189 | -0.192 | 0.003 | | | |
| z4 & z6 | | z4 | 0.3 | -0.234 | -0.259 | 0.024 | -0.024 | 0.170 | 0.829 |
| | | z6 | 0.3 | 0.288 | 0.290 | 0.002 | | | |
| | | z4 | 0.4 | -0.361 | -0.383 | 0.022 | -0.019 | 0.201 | 0.907 |
| | | z6 | 0.4 | 0.362 | 0.387 | 0.025 | | | |

Table 7: Summary of the laboratory measurements (Zernike terms & error characterization) results comparing 150 micron MLA with the 1.5 mm diameter polymer fibre bundle WFS (FBWFS) with the Thorlabs SHWFS. The last 4 columns are specified in microns.

| Aberration | | Zt | value | FBWFS | Th.SHWFS | $WF\Delta$ | medfit | RMS | PV |
|---|---|---|---|---|---|---|---|---|---|
| Tilt | Vert. | z2 | 0.3 | 0.3017 | 0.2569 | 0.0448 | 0.0280 | 0.1043 | 0.8068 |
| | | | 0.5 | 0.5227 | 0.4076 | 0.1151 | 0.0284 | 0.1574 | 1.3371 |
| | Horizt. | z3 | 0.3 | -0.2672 | -0.3000 | 0.0328 | -0.0062 | 0.0727 | 0.5813 |
| | | | 0.5 | -0.4213 | -0.4557 | 0.0344 | -0.0275 | 0.0905 | 0.6579 |
| Defocus | | z4 | 0.3 | 0.2589 | 0.2584 | 0.0005 | -0.0560 | 0.4292 | 1.6640 |
| | | | 0.5 | 0.4594 | 0.3468 | 0.1126 | -0.1307 | 0.6757 | 3.7285 |
| Astigmatism | Oblique | z5 | 0.3 | -0.1917 | -0.2303 | 0.0386 | -0.0279 | 0.1979 | 0.9663 |
| | | | 0.5 | -0.4719 | -0.4024 | 0.0695 | 0.0784 | 0.5222 | 3.0082 |
| | Vert. | z6 | 0.3 | 0.2393 | 0.2645 | 0.0252 | -0.0542 | 0.2945 | 1.331 |
| | | | 0.5 | 0.3823 | 0.3703 | 0.012 | -01975 | 0.416 | 1.907 |
| Coma | Vertical | z7 | 0.3 | -0.1130 | -0.1964 | 0.0834 | -0.492 | 0.480 | 2.948 |
| | | | 0.5 | -0.0138 | 0.0801 | 0.0663 | -0.459 | 0.371 | 2.162 |
| | Horizt. | z8 | 0.3 | -0.0603 | -0.0091 | 0.0512 | -0.116 | 0.302 | 1.569 |
| | | | 0.5 | -0.0828 | -0.0302 | 0.0526 | -0.513 | 0.476 | 0.476 |
| Trefoil | Vert. | z9 | 0.3 | -0.2284 | -0.1906 | 0.0378 | -0.037 | 0.267 | 1.145 |
| | | | 0.2 | -0.0992 | -0.0793 | 0.0199 | -0.031 | 0.158 | 0.959 |
| | Oblique | z10 | 0.3 | 0.0849 | 0.0838 | 0.0011 | 0.041 | 0.191 | 1.351 |
| z3,z5 & z7 | | z3 | 0.3 | 0.1195 | 0.0960 | 0.0235 | | | |
| | | z5 | 0.3 | 0.0822 | 0.0803 | 0.0019 | -0.0569 | 0.2327 | 1.3823 |
| | | z7 | 0.3 | -0.0387 | -0.0157 | 0.023 | | | |
| z4 & z6 | | z4 | 0.3 | -0.1296 | -0.1668 | 0.0372 | -0.0211 | 0.2042 | 1.3638 |
| | | z6 | 0.3 | 0.1951 | 0.1763 | 0.0188 | | | |
| | | z4 | 0.4 | -0.0968 | -0.1250 | 0.0282 | -0.0324 | 0.3258 | 1.5358 |
| | | z6 | 0.4 | 0.2273 | 0.2037 | 0.0236 | | | |

## 6. APPLICATIONS – MOAO

The science resulting from wide-field multi-object spectroscopy is an exciting prospect in the future era of the Extremely Large Telescopes (ELT), but Smart Focal Plane (SFP) devices are needed to enable maximum utilization of the focal plane surfaces. We believe that a promising candidate for such a SFP device for ELTs is a field-deployable mini-SHWFS positioned by Starbugs. The proposed MANIFEST fibre-positioner instrument for the Giant Magellan Telescope (GMT) is a possible application of AAO mini-SHWFS, see Figure 17 (a). The mini-SHWFS can facilitate as an upgrade option for MANIFEST with minimal cost. The mini-SHWFS could be fitted to a standard Starbug to enable wavefront sensor for either image quality diagnostics or for MOAO (partial corrections). The MANIFEST field plate diameter is 1.25 m or an equivalent field of view of 21 arcminutes. This would allow for a large number of mini-SHWFS to cover the field. A single wavefront corrector device (used in open-loop), can then be fed by a handful of mini-SHWFS to improve performance of the AO system; see Figure 17 (b).

The issue of the lack of sky coverage for the mini-SHWFS arising from the requirement of V <10 magnitude guide stars can be solved with the use of multiple laser guide stars (LGS), such as Rayleigh or Sodium. The Rayleigh LGS is often more powerful and more affordable than the Sodium LGS. The Ralyeigh beacon at a height of 10 km would be suitable as a probe of the turbulence up to a height of ~300 m above the GMT. Therefore, the mini-SHWFS using Raleigh LGS would be effective on nights that are dominated by the ground-layer turbulence. Turbulence above 300 m would not be sufficiently sampled with a Raleigh beacon due to the 'cone effect' and a Sodium LGS is required [1].

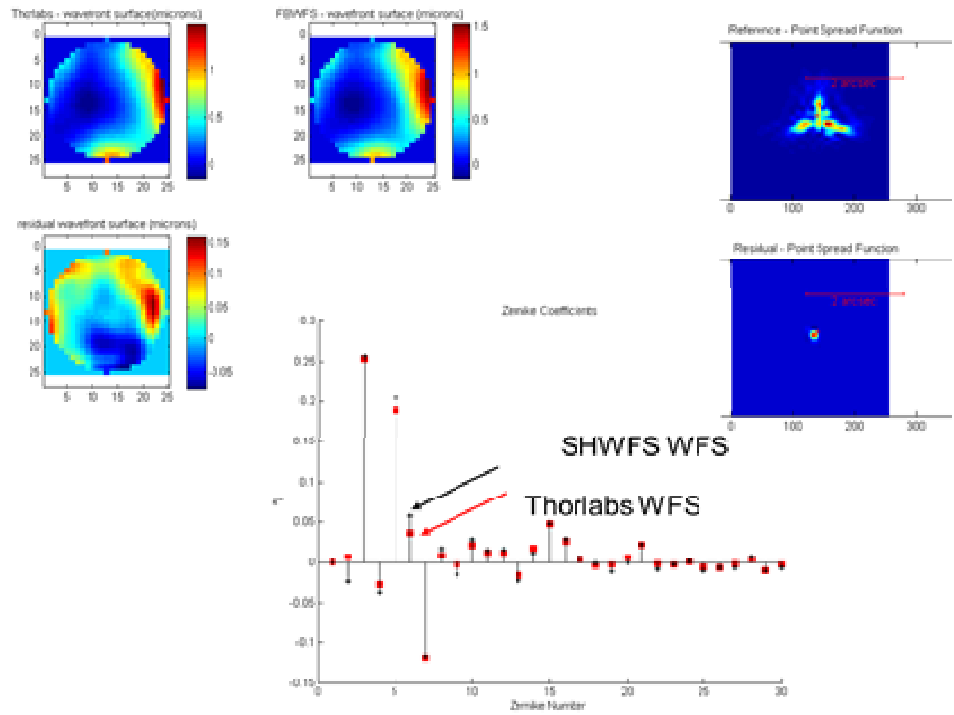

Figure 16: Example result showing good agreement with the reconstructed Zernike terms between the bundle mini-SHWFS (bundle SHWFS, black dots) and the Thorlabs SHWFS (red squares). In this case, the measured wavefront of the mini-SHWFS, if applied correctly to the deformable mirror, would achieve diffraction limited PSF.

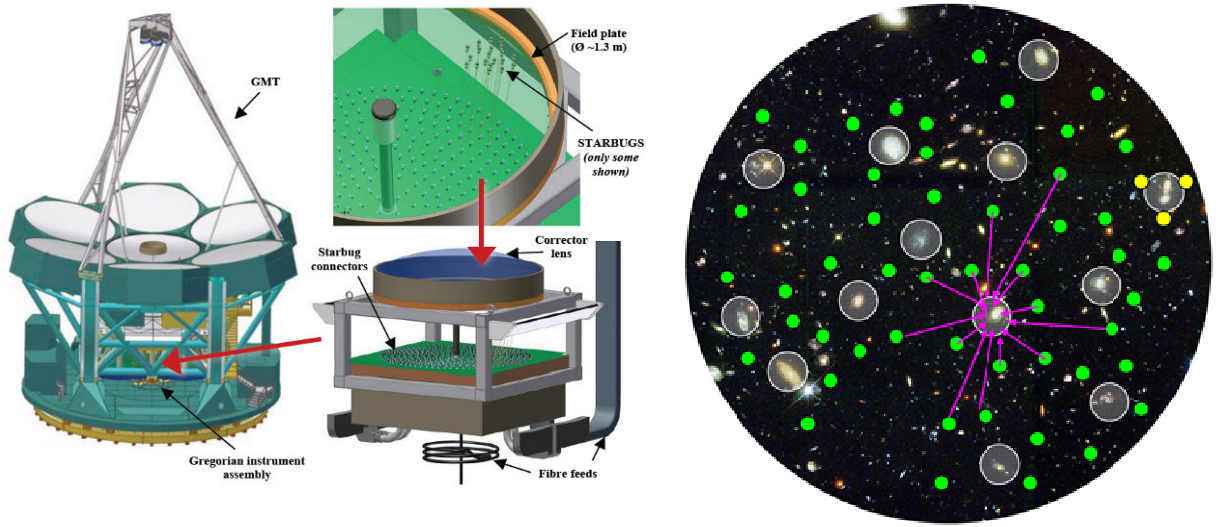

(a) GMT and MANIFEST (Starbugs positioner)　　　　(b) Starbugs mini-SHWFS for MOAO (green dots);

Figure 17: The mini-SHWFS can be used for MOAO applications, e.g. a possible upgrade option for MANIFEST.

## 7. CONCLUSIONS

The proposed mini-SHWFS further extends the versatility of Starbugs positioners. The ability to fit into a 'standard' Starbug minimizes both cost and technical risk. Therefore, the mini-SHWFS can leverage off the many advantages provided by Starbugs. The advantage to multiplex a relatively large number of mini-SHWFS onto a single low-noise fast-readout detector provides a substantial cost savings per wavefront sensor. The advantage of having multiple distributed WFS across the field provides exciting opportunities for new types of AO instruments.

The mini-SHWFS are not intended for precise wavefront measurements due artefacts caused by the bundle sampling and re-imaging optics. The mini-WFS are suitable to those applications requiring a large number of mid-to-low wavefront measurements for partial corrections, such as GLAO and MOAO.

We have proposed a concept design for a 13 x 13 mini-SHWFS system for the 3.9 m AAT at f/8. The system concept is capable of simultaneous operating a total of 66 mini-SHWFS at 198 fps (slow mode) or 9 mini-SHWFS at 419 fps (fast mode) for guide sources as faint as 10 mag. V-band (slow mode). Verification of the mini-SHWFS concept design has been shown through simulation and with laboratory measurements. Good agreement has been found between the polymer coherent imaging bundle based mini-SHWFS and the classical SHWFS. Further work for the mini-SHWFS includes developing a prototype for installation into a Starbug and working towards an on-sky demonstrator.